\documentclass[11pt]{article}
\usepackage{moriond,epsfig}

\def\beq{\begin{equation}}
\def\eeq{\end{equation}}
\def\beqa{\begin{eqnarray}}
\def\eeqa{\end{eqnarray}}

\def\lsim{\mbox{\raisebox{-.6ex}{~$\stackrel{<}{\sim}$~}}}
\def\gsim{\mbox{\raisebox{-.6ex}{~$\stackrel{>}{\sim}$~}}}
\def\pref#1{(\ref{#1})}

\begin{document}
\vspace*{4cm}
\title{EXTRA DIMENSIONS AND THE COSMOLOGICAL CONSTANT PROBLEM}

\author{ C.P. BURGESS }

\address{Department of Physics \& Astronomy, McMaster University,
Hamilton, ON, Canada\\
Perimeter Institute for Theoretical Physics, Waterloo, ON, Canada}

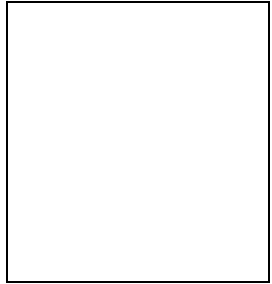
\maketitle\abstracts{This article reviews the arguments why extra
dimensions provide a unique opportunity for progress on the
cosmological constant problem, and updates the status of -- and
the objections to (with replies) -- the specific proposal using
supersymmetric large extra dimensions (SLED).}

\section{Extra Dimensions and the Cosmological Constant Problem}

For thirty years technical naturalness -- the requirement that
small parameters be stable against renormalization \cite{TechNat}
-- has been a major guideline for searches to replace the Standard
Model. For instance, the observation that particles with mass $M$
contribute to the Higgs potential an amount $\delta V_H \propto
M^2 H^* H$ leads to the Hierarchy Problem: how can $M_w / M_p \sim
10^{-15}$ be technically natural if any particles at all have
masses between $M_w$ and $M_p$? Naturalness would be assured if
the Higgs were composite at a scale $\Lambda_c \gsim M_w$ since
then there is no potential for heavy particles to correct above
the scale $\Lambda_c$. It would also be assured if the Higgs were
elementary but supersymmetry, broken at scales $\Lambda_s \gsim
M_w$, enforced the cancellation of bosons and fermions in their
contribution to $\delta V_H$. Such considerations significantly
shaped the design of the LHC, in order to test both of these
proposals.

\subsubsection{Naturalness in crisis}

Yet the discovery \cite{DEDisc} that the universe is now entering
an epoch of accelerated expansion has provoked an unprecedented
angst \cite{Weinberg,Anthropic} about the use of technical
naturalness as a fundamental theoretical criterion. It does so
because the acceleration is well described by adding a
cosmological constant, $\lambda$, to Einstein's equations,
\begin{equation} \label{EinsteinEq}
    R_{\mu\nu} - \frac12 \, R g_{\mu\nu} + \lambda \, g_{\mu\nu} =
    \frac{T_{\mu\nu}}{M_p^2} \,,
\end{equation}
with the required constant much smaller than most other
fundamental scales we know. Regarded as an energy density,
$\lambda = \rho/M_p^2$, observations require $\rho = \mu^4$, with
$\mu \lsim 10^{-2}$ eV.

Since particles of mass $M$ contribute $\delta \rho \propto M^4$,
the contribution of almost all known particles to $\rho$ are
already much too large: for electrons $m_e^4/\mu^4 \sim 10^{36}$,
for the QCD phase transition $\Lambda_{QCD}^4/\mu^4 \sim 10^{44}$,
while for electroweak bosons $M_w^4/\mu^4 \sim 10^{56}$. The
contributions of particles with $M \gg M_w$ are generically larger
still, but can be suppressed (such as by supersymmetry) to
contribute only $\delta \rho \sim M_w^4$.

This makes the cosmological constant (CC) problem the mother of
all naturalness problems, since its roots lie with particles we
already know rather than hypothetical particles having $M \gg
M_w$. With naturalness as their guide, theorists unaware of
accelerators operating above $10^{-2}$ eV might confidently
predict the discovery of supersymmetric partners split in mass
from the electon by this scale, in order to solve the CC problem.
How can we trust naturalness as a guide at the electroweak scale
if it lets us down so badly on scales we thought we understood?

\subsubsection{How extra dimensions can help}

The essence of the problem is that the Lorentz invariance of the
vacuum implies that the vacuum stress energy satisfies
$T^{vac}_{\mu\nu} = -\rho \, g_{\mu\nu}$, making $\rho/M_p^2$
indistinguishable from $\lambda$ in eq.~\pref{EinsteinEq}. The
puzzle is how the curvature of space (and so also $\lambda$) can
be as small as is measured when quantum corrections to $\rho$
should be large.

Extra dimensions help by breaking the link between 4D Lorentz
invariant energies ($\rho$) and 4D curvature ($\lambda$). They can
do so because if the vacuum energy is associated with the tension
of a surface (or brane), then it is localized (and not Lorentz
invariant) in the extra dimensions. Although it necessarily curves
spacetime, it sometimes does so by curving the extra dimensions
and not the four dimensions we see.\cite{CLP,5DSelfTune,SLED1}

This is all very well, but any extra dimensional model becomes
effectively four dimensional at energies below its Kaluza-Klein
scale, $\Lambda_{\scriptscriptstyle KK} \sim 1/r$, where $r$
denotes a generic linear size (radius) for the largest extra
dimensions. Consequently an intrinsically extra-dimensional
explanation of the size of $\rho$ can only be useful if the extra
dimensions are large: $\Lambda_{\scriptscriptstyle KK}$ cannot be
too much larger than $\mu \sim 10^{-2}$ eV, so $r$ can't be much
smaller than $\sim 10$ $\mu$m. Remarkably, extra dimensions can
actually be this large,\cite{ADD} but only within a `brane-world'
scenario for which all observed particles are trapped on a 4D
surface (or 3-brane). In this case only gravitational measurements
probe the extra dimensions, and constraints on deviations from
Newton's Law presently allow dimensions slightly smaller than 50
$\mu$m.\cite{NLTests} Most encouragingly, large extra dimensions
potentially do just what one wants: because observed particles are
trapped on a brane, their non-gravitational properties are
unchanged (as they must be) at the energies to which we have
access. All that is modified is how their their vacuum energy
gravitates.

The extra-dimensional approach to the CC problem starts with this
observation, and asks whether the theoretical elbow room thus
opened is sufficient to allow a small enough 4D curvature in a
technically natural way. This involves re-asking the cosmological
constant problem in higher dimensions: What choices are required
to make our observed 4 dimensions very flat? And can these choices
be stable against renormalization? So far these issues are most
thoroughly explored in 6 dimensions, to which we now turn.

\section{Supersymmetric Large Extra Dimensions (SLED)}

The best-developed proposal along these lines is the SLED
proposal,\cite{SLED1,SLED2,SLEDRev,TAMU} according to which all
known particles are localized on one of possibly many (usually
two) parallel 3-branes that are situated at points within a 6D
spacetime whose two extra dimensions are at present imagined to be
$r \sim 10$ $\mu$-metres in size, so that $1/r \sim 10^{-2}$ eV is
not so different from $\mu$. It is further assumed that the `bulk'
physics -- not trapped on the branes -- is supersymmetric, and so
is described by any one of the many known 6D supergravities. If
the extra dimensions are not too strongly warped (as is true for
the majority of explicit solutions known \cite{GGP,SLED2}) the 4D
Planck mass is of order \cite{ADD} $M_p \sim M_g^2 r$, so the
scale of the 6D Newton constant must be $M_g \sim 10$ TeV. The
bulk supersymmetry is imagined to be badly broken by the branes,
whose tension is imagined to be of the order of (but somewhat
smaller than) $M_g$.

This proposal is the best developed in several senses. First, it
is the one for which the naturalness issues have been the most
thoroughly
explored.\cite{SLEDUV,SLEDdS,SLEDtdep,Linearized,Hypers} Second,
it is (so far) the only extra-dimensional framework that does not
argue for a vanishing 4D curvature, but instead provides an
explicit mechanism for a nonzero 4D curvature of size $\mu$.
Finally, it leads to a known low-energy 4D field theory within
which gravity is described by a scalar-tensor system, with the
scalar labelling the classical flat direction corresponding to
overall re-scalings of the extra dimensions, within which a
realistic quintessence-type accelerated expansion can plausibly
take place.\cite{SLEDCosmo,Kimmo,Cosmo2}

Best of all, the extra dimensions themselves must be very large,
$1/r \sim \mu$, and the scale of gravity in the extra dimensions
must be low, $M_g \sim 10$ TeV. As a result the proposal is
unusually predictive --- with many testable predictions for tests
of gravity and for particle colliders,\cite{SLEDPheno} in addition
to its implications for cosmology.

\subsection{Where we stand}

The SLED proposal involves re-asking the CC problem in higher
dimensions. This comes in two steps: ($i$) enumerate the choices
which are required to obtain a small 4D curvature within a
particular extra-dimensional context; ($ii$) identify whether or
not these choices are stable against renormalization (and so are
technically natural).

\subsubsection{What is required for 4D flatness?}

Most of the progress so far has been in identifying what choices
are required for matter on the branes in order to obtain
compactifications whose 4D geometry is approximately flat.
Although it is not crucial for the naturalness arguments, these
choices are best explored within chiral gauged 6D
supergravity.\cite{NS} Although not the simplest, this
supergravity receives special attention because it allows
spherical compactifications, and so can involve only
positive-tension branes.\cite{SS,SLED1,SLED2,GGP} Spherical extra
dimensions are related to positive-tension branes by a topological
argument, which is easiest to see for branes whose tension, $T_b$,
back-reacts on the geometry to give a conical defect, with defect
angle $\delta_b = T_b/M_g^4$ (so $\delta_b/2\pi = 4GT_b$). The
Euler number, $\chi$, for the extra dimensions then is
\begin{equation} \label{Euler}
    \chi = 4G \sum_b T_b
    + \frac{1}{4\pi} \int d^2 x \; \sqrt{g} \, R_2
    \,,
\end{equation}
where $R_2$ is the 2D geometry's Ricci scalar. Notice that for
toroidal compactifications $R_2 = \chi = 0$, so the brane tensions
must all sum to zero (and some in particular must be negative). On
the other hand, for spherical compactifications $\chi = 2$, so all
of the tensions can be positive.

A broad class of exact solutions to 6D gauged, chiral supergravity
are now known, including those which are 4D
flat,\cite{SLED1,SLED2,GGP,Hypers} those having curved 4D maximal
symmetry\cite{SLEDdS} and those which are
time-dependent.\cite{SLEDtdep} These show that solutions
appropriate to two source branes are generically time-dependent,
describing geometries wherein the extra dimensions implode or run
away to flat 6D space. It turns out that for codimension-two
branes a sufficient condition that ensures that all static
solutions are 4D flat is to have the branes not couple to the 6D
dilaton (a scalar which is partnered to the graviton by 6D
supersymmetry).\cite{SLED2,SLEDtdep}

\subsubsection{How stable are these choices?}

The next question asks how stable are the choices required to make
the observed 4 dimensions flat. This question comes in two
separate parts: ($i$) are the choices required for couplings in
the action stable against renormalization; and ($ii$) given
specific choices for the action, are acceptable solutions stable
against changes to the initial conditions.

\medskip\noindent{\em Stability to initial conditions:} Given the
number of solutions now known it is clear that, even given
appropriate brane properties, the generic solutions to 6D
supergravity describe time-dependent runaways.\cite{SLEDtdep} This
shows that extra dimensional approaches to the CC problem
generically have an initial condition problem: they describe the
universe around us only if the universe starts out in a particular
way. This makes them like the Hot Big Bang model itself, whose
similar initial-condition problems inspired the invention of
inflationary scenarios. Since the plausibility of initial
conditions for the later universe can potentially be addressed by
changing the dynamics of the earlier universe (such as through
inflation), this kind of initial-condition problem is a price
worth paying if it allows progress to be made on the more
difficult issue of technical naturalness.

\medskip\noindent{\em Stability to renormalization:} The key
question is whether the choices which allow 4D flat solutions are
natural, in the sense of being stable against renormalization.
Once arranged as desired, do these choices stay made as heavy
particles are integrated out? Although work along these lines is
still in progress, some partial results are known.\cite{SLEDUV}

It is known that the Casimir energy produced by integrating out
bulk fields for a toroidal bulk have the desired size. More
generally, integrating out heavy bulk particles at one loop tends
not to cause problems because these loops know about the full 6D
supersymmetry of the action. It is the 6D supersymmetry which is
relevant to integrating out the dangerous frequencies, $\omega
\gsim M_w$, even though 6D SUSY is broken by the background
geometry. This is because these dangerous modes probe very short
distances in the bulk, and so are largely insensitive to the
geometry over scales $\sim r$ (as they would have to be to `know'
that supersymmetry breaks).

In some circumstances integrating out massive brane fields also
need not be dangerous, despite supersymmetry being badly broken on
the branes. This is because a sufficient condition for 4D flatness
is the absence of a brane coupling to the bulk dilaton, and
arbitrary numbers of brane loops cannot generate a coupling to the
dilaton if it is not already present at the classical level.

The potentially most dangerous contributions are those which mix
brane and bulk loops, since these can introduce couplings between
the brane and the dilaton and can know about supersymmetry
breaking. The good news here is that it is sufficient to establish
that these contributions are small to a small number of loops in
the bulk, because the very large size of the extra dimension
implies the bulk loops cost a factor of order $1/r^2$. (Recall
that the observations require a 4D energy density of order $\mu^4
\sim 1/r^4$.) It is these calculations of naturalness on which the
success or failure of the SLED proposal must ultimately be judged.

\section{Some Objections}

It is useful to close by listing some of the best objections which
have been raised against the SLED proposal over the years,
together with a cartoon of the arguments as to why they do not
(yet) appear to be show-stoppers.

\subsubsection{Why isn't SLED killed by the arguments against 5D
self-tuning?}

The observation that higher dimensions can allow 4D flat geometry
to coexist with nonzero brane tension was explored
\cite{5DSelfTune} and rejected \cite{5DSTx} within a 5D context.
Given the similarity in their motivations it is natural to wonder
if the 6D proposal can be killed in a similar way.

The objection in the 5D case argued that hidden fine-tunings were
involved, because the presence of a brane with positive tension,
$T_1$, necessarily warps the transverse dimensions and forces the
bulk geometry to have a singularity elsewhere in the bulk. This
singularity is naturally interpreted as the presence of a second
brane, and on general grounds this second brane is found to have a
negative tension, $T_2$, with 4D flatness requiring a cancellation
between $T_1$ and $T_2$. How could this cancellation possibly
survive integrating out heavy physics on only one brane?

The analogue of this 5D argument arises for 4D-flat
compactifications of 6D supergravity on a torus. In this case
using $R_2 = \chi = 0$ in eq.~\pref{Euler} implies $\sum_b T_b =
0$, which shows that all such compactifications require
cancellation amongst brane tensions. But crucially
eq.~\pref{Euler} does {\it not} rely on 4D flatness. Rather, being
topological it must continue to hold under any continuous
perturbation, such as the `integrating out' of short-wavelength
physics. If, in particular, the tension is adjusted on only one
brane then eq.~\pref{Euler} remains true, and implies the bulk
geometry must necessarily curve in response. A real calculation is
required to see whether the observed 4 dimensions also curve.

\subsubsection{What about Weinberg's Theorem?}

Weinberg has a general no-go theorem \cite{Weinberg} against
approaches to the CC problem (like SLED) which rely on
spontaneously broken classical scale invariance. In a nutshell,
this states that scale invariance by itself cannot protect the CC
from quantum corrections, even in the absence of scale anomalies.
As discussed in more detail elsewhere,\cite{TAMU} as applied to
SLED Weinberg's argument correctly implies that the
scale-invariant classical flat direction of 6D supergravity must
be lifted by quantum corrections. It does not in itself say {\it
how big} these corrections must be. In SLED this lifting is
partially protected by the unusually small bulk-supersymmetry
breaking scale, $\Delta m_s^2 \sim 1/r^2$, and so must vanish as
$r \to \infty$. The key work in progress for SLED is showing that
this suppression is of order $1/r^4$ (in the Jordan frame) rather
than merely being of order $M_w^4$ or $M_w^2/r^2$. (Notice that
although $M_w^2/r^2$ is too large to be consistent with the
observed Dark Energy, it is parameterically small compared with
the normal size, $O(M_w^4)$, usually obtained within models.)

\subsubsection{What is the 4D relaxation mechanism for Self-Tuning?}

This objection argues against the possibility of `self-tuning',
{\it i.e.} of there being any system for which perturbations of
initially 4D-flat solutions lead to new static 4D flat solutions.
It arises in the following two different forms: \footnote{We thank
N. Arkani-Hamed, G.Dvali and J.Polchinski for arguments along
these lines.}

\medskip\noindent {\it Volume Stabilization and Self-Tuning:}
One form of the argument argues that eventually it is necessary to
stabilize the extra dimensions, and so develop a minimum for some
sufficiently large value of $r \sim r_0$, with an effective 4D
potential satisfying $V(r_0) \lsim 1/r_0^4$. Once this is done,
self-tuning would require the potential to adjust dynamically to
any changes of microscopic parameters (such as a phase transition
on one of the branes) in such a way as to obtain a new stabilized
minimum at $r \sim r_0'$, with $r_0' \sim r_0$ and $V(r_0') \lsim
1/r_0^4$. But there is no known way to do this within a 4D
effective description.

While this seems to be a true statement, SLED is {\it not} a
self-tuning system. At the classical level this can be seen
because of its flat direction: any classical perturbation
stimulates a roll along the flat direction and so is not attracted
towards a new static solution. Furthermore, this property can
survive the lifting of the flat direction by quantum corrections,
because the resulting $1/r^4$ potential typically does not
stabilize the volume, and also favours a runaway along the
would-be flat direction. At the 4D level the low-energy dynamics
is described by a scalar-tensor theory, with the $1/r^4$ potential
naturally giving a scalar mass of order the Hubble scale: $m \sim
H \sim \mu^2/M_p$.

What is remarkable is that a light scalar with a potential of the
form obtained by dimensional reduction can potentially provide a
description of what is seen in cosmology. To see how this might
work, imagine that the classical flat direction along which $r$
can change (parameterizing the classical scale invariance of the
6D equations) is lifted by quantum effects that are of order
$1/r^4$ (which of course is the hard part, see above). Such a
potential drives a runaway out to $r \to \infty$, without
stabilizing at any fixed $r$. However, the full quantum
contributions to the potential generically also include
logarithms: $V(r) \sim (1/r^4)[a + b\log r + \cdots]$. It happens
that this kind of potential can describe a successful
quintessence-type cosmology,\cite{AS} with potential domination
occurring when \cite{SLEDCosmo} $\log r \sim a/b$. Of course the
success of this cosmology requires finding a compactification for
which $a/b \sim 60$ (which is not yet done); that the universe
starts off with somewhat special initial conditions (which we
expect in any case from the 6D point of view); and that the light
scalar which results is not ruled out by tests of gravity (also
possible -- but not generic -- given the $\log r$ corrections to
its matter couplings \cite{SLEDCosmo}). But these are all issues
which are likely to be easier to solve than is the original
cosmological constant problem. \footnote{An alternative
possibility \cite{Kimmo} has the coefficient, $A(r)$, of the
kinetic term, $A(r) (\partial r)^2$, vanishing for $r \sim r_0$.}

\medskip\noindent {\it Adiabatic Version:} An alternative
version of this argument starts from the observation that there is
nothing in a low-energy effective theory which precludes having a
CC which is larger than the cutoff, provided that it is turned on
in a sufficiently adiabatic way. As applied to extra dimensional
models, this appears to mean that a CC larger than $1/r^4$ could
make sense in the effective 4D theory designed to describe physics
at scales $E \ll 1/r$. This argues that extra dimensions {\it
cannot} be key to the argument, since it should be possible to
understand purely in four dimensions why one cannot add a large CC
compared with the present-day Kaluza-Klein (KK) scale.

The loophole in this argument lies in its ignoring of the scale
invariance of the higher-dimensional models, which preclude having
a strictly constant term in the low-energy potential. Rather, any
large energy density must really arise in the low-energy theory as
a potential for $r$, of the form $M^{4-n}/r^n$. However, for
canonically normalized fields, $\omega \sim M_p\log (Mr)$, this
becomes $A e^{\lambda \omega/M_p}$, for some $\lambda$, with $A
\sim M^4$ large. However, besides providing a large energy
density, such a term also acts as a force pushing $\omega$,
implying in particular $\dot\omega^2 \sim M^4$. Since this implies
$\dot\omega$ is greater than the KK scale, it is necessarily
non-adiabatic and so not describable purely within the 4D theory.

\section{Summary}

Applying extra dimensions to the cosmological constant problem is
clearly a work in progress. However the stakes are high and
include the validity of naturalness as a fundamental theoretical
criterion. On the one hand extra dimensions are attractive,
inasmuch as they break the basic link between vacuum energy and 4D
curvature which is at the root of the CC problem. On the other
hand, it is not yet known whether loop corrections can be as small
as a truly natural solution to the CC problem would require
(although this should be known very soon).

Moreover, even if extra dimensions provide stability under
renormalization, they inevitably appear to involve special choices
of initial conditions if they are to describe our present-day
universe. Whether this is a reasonable trade-off, or involves
throwing out the baby with the bathwater, can only be decided by
examining extra dimensional solutions in more detail.

\section*{Acknowledgements}

This work was supported in part by funds from NSERC (Canada), the
Killam Foundation and McMaster University, and summarizes the
results of collaborations with Yashar Aghababaie, Jim Cline,
Claudia de Rham, Hassan Firouzjahi, Doug Hoover, Quim Matias,
Susha Parameswaran, Fernando Quevedo, Gianmassimo Tasinato, Andrew
Tolley and Ivonne Zavala. I'm indebted to the organizers of the
Moriond Electroweak Workshop for providing the invitation, setting
and intellectual milieu that make their meetings unique.


\begin{thebibliography}{99}

\bibitem{TechNat}
G. 't Hooft, in {\sl Cargese Summer Inst.} 1979:135
(QCD161:S77:1979) (reprinted in {\sl 't Hooft, G. (ed.): Under the
spell of the gauge principle} 352-374, and in {\sl Farhi, E.
(ed.), Jackiw, R. (ed.): Dynamical gauge symmetry breaking}
345-367).

\bibitem{DEDisc}
S.~Perlmutter {\it et al.}, Ap.\ J.\ {\bf 483}, 565 (1997),
[astro-ph/9712212];
%
A.G. Riess {\it et al.}, Ast.\ J.\ {\bf 116}, 1009 (1997),
[astro-ph/9805201];
%
N. Bahcall, J.P. Ostriker, S. Perlmutter, P.J. Steinhardt, Science
{\bf 284}, 1481 (1999), [astro-ph/9906463].

\bibitem{Weinberg}
S. Weinberg, {\it Rev. Mod. Phys.} {\bf 61} (1989) 1.

\bibitem{Anthropic}
R.~Bousso and J.~Polchinski,
  JHEP {\bf 0006} (2000) 006
  [hep-th/0004134];
%
L. Susskind,
[hep-th/0302219];
%
 J.~Polchinski,
  [hep-th/0603249].

\bibitem{CLP}
J.-W. Chen, M.A. Luty and E. Pont\'on, JHEP 0009 (2000) 012,
[hep-th/0003067];
%
S.M. Carroll and M.M. Guica, [hep-th/0302067];
%
I. Navarro, JCAP 0309:004,2003, [hep-th/0302129].

\bibitem{5DSelfTune}
N.~Arkani-Hamed, S.~Dimopoulos, N.~Kaloper and R.~Sundrum,
Phys.\ Lett.\ B {\bf 480} (2000) 193, [hep-th/0001197];
%
S.~Kachru, M.~B.~Schulz and E.~Silverstein,
Phys.\ Rev.\ D {\bf 62} (2000) 045021, [hep-th/0001206].

\bibitem{SLED1}
Y. Aghababaie, C.P. Burgess, S. Parameswaran and F. Quevedo,
Nucl.\ Phys.\ {\bf B680} (2004) 389--414, [hep-th/0304256].

\bibitem{ADD}
N. Arkani-Hamed, S. Dimopoulos and G. Dvali, { Phys.\ Lett.} {\bf
B429} (1998) 263  hep-ph/9803315; Phys.\ Rev.\ {\bf D59} (1999)
086004  [hep-ph/9807344];
%
I.~Antoniadis, N.~Arkani-Hamed, S.~Dimopoulos and G.~R.~Dvali,
Phys.\ Lett.\ B {\bf 436} (1998) 257 [hep-ph/9804398].

\bibitem{NLTests}
For a review with references, see E.G. Adelberger, B.R. Heckel and
A.E. Nelson, Ann.\ Rev.\ Nucl.\ Part.\ Sci.\ {\bf 53} (2003)
77--121, [hep-ph/0307284].

\bibitem{SLED2}
Y. Aghababie, C.P. Burgess, J.M. Cline, H. Firouzjahi, F. Quevedo,
G. Tasinato and I. Zavala, JHEP 0309 (2003) 037 (48 pages)
[hep-th/0308064].

\bibitem{SLEDRev}
C.P. Burgess,
{\it Ann. Phys.} {\bf 313} (2004) 283-401 [hep-th/0402200].

\bibitem{TAMU}
  C.~P.~Burgess,
AIP Conf.\ Proc.\  {\bf 743} (2005) 417 [hep-th/0411140].

\bibitem{GGP}
G.W.~Gibbons, R.~G\"uven and C.N.~Pope,
Phys.\ Lett.\ B {\bf 595}, 498 (2004) [hep-th/0307238].

\bibitem{SLEDUV}
 C.~P.~Burgess and D.~Hoover,
  [hep-th/0504004];
  %
   D.~M.~Ghilencea, D.~Hoover, C.~P.~Burgess and F.~Quevedo,
  JHEP {\bf 0509}, 050 (2005)
  [hep-th/0506164];
  %
  D.~Hoover and C.~P.~Burgess,
  JHEP {\bf 0601}, 058 (2006)
  [hep-th/0507293];
  %
  E.~Elizalde, M.~Minamitsuji and W.~Naylor,
  Phys.\ Rev.\  D {\bf 75} (2007) 064032
  [hep-th/0702098];
  %
  C.P.~Burgess, D.~Hoover and G.~Tasinato,
  arXiv:0705.3212 [hep-th].

\bibitem{SLEDdS}
  A.J.~Tolley, C.P.~Burgess, D.~Hoover and Y.~Aghababaie,
   JHEP {\bf 0603} (2006) 091 [hep-th/0512218].

\bibitem{SLEDtdep}
  A.J.~Tolley, C.P.~Burgess, C.~de Rham and D.~Hoover,
  New J.\ Phys.\  {\bf 8} (2006) 324
  [hep-th/0608083];
  %
  CITE Ed Copeland's new solutions.

\bibitem{Linearized}
    J.M.~Cline, J.~Descheneau, M.~Giovannini and J.~Vinet, JHEP 0306
    (2003) 048 [hep-th/0304147];
%
  H.~M.~Lee and A.~Papazoglou,
  [hep-th/0602208];
%
  C.~P.~Burgess, C.~de Rham, D.~Hoover, D.~Mason and A.~J.~Tolley,
  JCAP {\bf 0702} (2007) 009
  [hep-th/0610078].

\bibitem{Hypers}
%
   S.~Randjbar-Daemi and E.~Sezgin,
  Nucl.\ Phys.\ B {\bf 692} (2004) 346
  [hep-th/0402217];
%
 A.~Kehagias,
  Phys.\ Lett.\ B {\bf 600} (2004) 133
  [hep-th/0406025];
%
  S.~Randjbar-Daemi and V.~A.~Rubakov,
  JHEP {\bf 0410}, 054 (2004)
  [hep-th/0407176];
%
  H.~M.~Lee and A.~Papazoglou,
  Nucl.\ Phys.\ B {\bf 705} (2005) 152
  [hep-th/0407208];
%
 V.~P.~Nair and S.~Randjbar-Daemi,
  JHEP {\bf 0503} (2005) 049
  [hep-th/0408063];
%
   S.~L.~Parameswaran, G.~Tasinato and I.~Zavala,
  [hep-th/0509061];
%
 H.~M.~Lee and C.~Ludeling,
  [hep-th/0510026].

\bibitem{SLEDCosmo}
    A. Albrecht, C.P. Burgess, F. Ravndal and C. Skordis,
    Phys. Rev. {\bf D65} (2002) 123507 [astro-ph/0107573];

\bibitem{Kimmo}
  K.~Kainulainen and D.~Sunhede,
  Phys.\ Rev.\  D {\bf 73} (2006) 083510
  [astro-ph/0412609].

\bibitem{Cosmo2}
  L.~Anchordoqui, H.~Goldberg, S.~Nawata and C.~Nunez,
  arXiv:0704.0928 [hep-ph].

\bibitem{SLEDPheno}
    J.~Matias and C.~P.~Burgess,
  JHEP {\bf 0509} (2005) 052
  [hep-ph/0508156];
  %
   P.~Callin and C.~P.~Burgess,
  [hep-ph/0511216].

\bibitem{NS}
H. Nishino and E. Sezgin, {\it Phys. Lett.} {\bf 144B} (1984) 187;
Nucl.\ Phys.\ {\bf B278} (1986) 353.

\bibitem{SS}
A.~Salam and E.~Sezgin,
Phys.\ Lett.\ B {\bf 147} (1984) 47;
%
S. Randjbar-Daemi, A. Salam, E. Sezgin and J. Strathdee, {\it
Phys. Lett.} {\bf B151} (1985) 351.

\bibitem{5DSTx}
S.~Forste, Z.~Lalak, S.~Lavignac and H.~P.~Nilles,
Phys.\ Lett.\ B {\bf 481} (2000) 360, hep-th/0002164; JHEP {\bf
0009} (2000) 034, [hep-th/0006139];
%
C.~Csaki, J.~Erlich, C.~Grojean and T.J.~Hollowood,
Nucl.\ Phys.\ {\bf B584} (2000) 359-386, [hep-th/0004133];
%
C.~Csaki, J.~Erlich and C.~Grojean,
Nucl.\ Phys.\ {\bf B604} (2001) 312-342, [hep-th/0012143];
%
J.M. Cline and H. Firouzjahi,
Phys.\ Rev.\ {\bf D65} (2002) 043501, [hep-th/0107198].

\bibitem{AS}
A. Albrecht and C. Skordis, {\it Phys. Rev. Lett.} {\bf 84} 2076
(2000) (astro-ph/9908085).

\end{thebibliography}
\end{document}